\documentclass[preprint,11pt,nofootinbib]{revtex4}
\usepackage{graphicx}% Include figure files
\usepackage{amsmath, amssymb, graphics, epsfig}
\usepackage{dcolumn}% Align table columns on decimal point
\usepackage{hyperref}
\usepackage{epstopdf}
\newcommand{\md}{\mathrm{d}}

\newcommand{\nc}{\newcommand}
\nc{\ba}{\begin{eqnarray}}
\nc{\ea}{\end{eqnarray}}
\newcommand\be{\begin{equation}}
\newcommand\ee{\end{equation}}
\nc{\K}{{\bf k }}
\nc{\x}{{\bf x }}
\nc{\D}{\overline{\mbox{D3}}}

\preprint{IPM/P-2011/006}

\begin{document}

\title{DBI Lifshitz Inflation}
\author{Mohsen Alishahiha}
\email{alishah(AT)ipm.ir}
\author{Hassan Firouzjahi}
\email{firouz(AT)ipm.ir}
\author{Mohammad Hossein Namjoo}
\email{mh.namjoo(AT)ipm.ir}

\affiliation{ School of Physics, Institute for Research in Fundamental Sciences (IPM),
P. O. Box 19395-5531,
Tehran, Iran}
%\date{\today}

\begin{abstract}

A new model of DBI inflation is introduced where the mobile brane, the inflaton field, is moving relativistically inside a  Lifshitz throat with an arbitrary anisotropic scaling exponent $z$. After dimensional reduction to four dimension the general covariance is broken explicitly both in the matter and the gravitational sectors. The general action for the metric and matter field perturbations are obtained and it is shown to be similar to the classifications made in the effective field theory of inflation literature. 

\end{abstract}
  
\maketitle
%%%%%%%%%%%%%%%%%%%%%%%%%%%%%%%%%%%%%%%%%%%%%%%%%%
\section{Introduction  }
\label{intro}

Cosmic Inflation \cite{Guth:1980zm, Linde:1981mu}
has emerged as a leading theory for the early universe and structure formation which is strongly supported by recent observations \cite{Komatsu:2010fb}. In simple models of inflation, the inflaton field is minimally coupled to gravity with potential flat enough to obtain a long enough 
 period of inflation to solve the flatness and the horizon problems. Simple models of inflation predict almost scale invariant, almost Gaussian and almost adiabatic curvature perturbations which are in very good agreement with observations. However, with the advance of data in coming years, it is expected that a small but non-zero amount of non-Gaussianities can be detected. This can be used to rule out simple models of inflation such as single field chaotic  scenarios. 
 
During last decade there have been extensive efforts to embed inflation in string theory, for a review see  \cite{HenryTye:2006uv,Cline:2006hu,Burgess:2007pz,McAllister:2007bg,Baumann:2009ni, Mazumdar:2010sa}.  Brane inflation is an interesting realization of inflation from string theory \cite{dvali-tye,Alexander:2001ks,collection,Dvali:2001fw}. In its original form, the scenario
contained a pair of D3  and anti D3 branes moving in the Calabi-Yau (CY) compactification. The inflaton field is the radial distance between the pair so in this sense the inflaton field has a geometric interpretation in string theory. Inflation ends when the distance between the brane and anti-brane reaches the string length scale where a tachyon develops in the open strings spectrum stretched between the pair. Inflation ends soon after tachyon formation and the energy stored in branes tensions are released into closed string modes  \cite{HenryTye:2006uv}. However, it was soon realized that the 
potential between the pair of brane and anti-brane is too steep to allow a long enough period of 
slow-roll inflation. To flatten the potential, it was suggested to put the pair of brane and anti-brane  inside a warped throat \cite{Klebanov:2000hb, Giddings:2001yu, Dasgupta:1999ss}, where the potential between D3 and $\D$ is warped down as in Randall-Sundrum scenario
\cite{Randall:1999ee, Kachru:2003sx, Firouzjahi:2003zy, Burgess:2004kv, Buchel, Iizuka:2004ct, Firouzjahi:2005dh}. Dirac-Born-Infeld (DBI) inflation \cite{Silverstein:2003hf, Alishahiha:2004eh} (see also \cite{Kehagias:1999vr}) is a specific model of brane inflation where the mobile brane, the inflaton field, is moving ultra relativistically inside an AdS throat. A novel feature of DBI inflation is the production of large non-Gaussianities which has significant observational 
implications to constrain the model parameters. This has attracted considerable interests in literature, for a review of DBI inflation and its implications for non-Gaussianities see 
\cite{Chen:2006nt, Chen:2004gc, Shandera:2006ax, Bean:2007eh, Bean:2007hc} and the references therein. 

In this work we extend the idea of DBI inflation to the background which is not  an AdS
throat. More precisely we will consider the case where the mobile brane is moving in a 
Lifshitz background. This geometry has attracted considerable attentions recently
in the context of non-relativistic AdS/CFT correspondence where it may 
provide a gravity description for Lifshitz fixed point. We note that Lifshitz fixed
points appear when we are dealing with a physical system at critical point
with anisotropic scale invariance. 

Actually  at critical points the physics is usually described by scale invariant 
phenomena. Typically
the scale invariance arises in the conformal group where we have
\be
t\rightarrow \lambda t,\;\;\;\;\;\;x_i\rightarrow \lambda x_i.
\ee
Here $t$ is time and $x_i$'s are spatial directions of the spacetime.   

We note, however, that in many physical systems the critical points are governed by
dynamical scaling in which the space and time scale differently. In fact  spatially 
isotropic scale invariance is characterized by the dynamical exponent $z$
 as follows \cite{Hertz:1976zz}
\be\label{Lif}
t\rightarrow \lambda^zt,\;\;\;\;\;\;x_i\rightarrow \lambda x_i.
\ee
The corresponding critical points are known as Lifshitz fixed points.

In light of AdS/CFT correspondence \cite{Maldacena:1997re} it is natural to seek for
gravity duals of Lifshitz fixed points. Indeed the gravity descriptions of  
Lifshitz fixed points have been first considered in \cite{Kachru:2008yh} where
a metric  invariant under the scaling \eqref{Lif} where introduced.
The corresponding geometry is given by\footnote{Lifshitz metric typically is a solution
of a gravitational theory coupled to gauge fields \cite{Kachru:2008yh}. Lifshitz metric may also be
a solution of pure gravity modified by curvature squared terms \cite{AyonBeato:2010tm}.}
\be\label{BacLif}
d s^2=-\left(\dfrac{r}{L} \right)^{2z} \md t^2+ \left(\dfrac{r}{L} \right)^2 \md \x^2+\left(\dfrac{L}{r} \right)^2 \md r^2
%ds^2=L^2\left(-r^{2z} dt^2+r^2d\vec{x}^2+\frac{dr^2}{r^2}\right),
\ee
where $L$ is  the curvature radius of the metric and $r$ is the radial coordinate of the 
``Lifshitz throat''. The action of the scale transformation \eqref{Lif} on the metric is given by
\be
\label{scaling}
t\rightarrow \lambda^zt,\;\;\;\;\;\;x_i\rightarrow \lambda x_i,\;\;\;\;\;\;
r\rightarrow \lambda^{-1}r.
\ee
As it has been mentioned in \cite{Kachru:2008yh} although the metric is nonsingular, 
it is not geodesically complete and in particular an in-falling object into $r=0$  feels
a large tidal force.

Since at $z=1$ the metric reduces to that of  AdS, it is interesting to see how  
physical models are 
affected when we replace an AdS geometry with Lifshitz metric. In particulate in this paper 
we would like to study this effect in a cosmological model. Indeed fast moving branes in an 
AdS throat may lead to interesting distinctive inflationary models  \cite{Silverstein:2003hf, Alishahiha:2004eh}.  Therefore
it is natural to look for inflationary models where the brane moves inside a Lifshitz throat.
Since in this case the background has a free parameter, $z$, one may find interesting
observational predictions when we vary $z$.

We note, however, that unlike AdS geometry which can be easily obtained in the
context of string theory\footnote{An AdS throat may be obtained in type II B string theory
compactified on a Calabi-Yau 3-fold by putting large enough D3 branes on top of each other 
at a point on the Calabi-Yau. Of course the throats arising from IIB flux compactification
 are not AdS at all length scales, but nevertheless can look approximately AdS in some 
scale of energy.}, it is far from obvious how to  embed a Lifshitz geometry in string theory.
Nonetheless, it is worth mentioning  that string theory realizations of Lifshitz geometries 
have been studied in \cite{Hartnoll:2009ns}  in the context of strange metallic holography,
see also \cite{Balasubramanian:2010uk, Donos:2010tu, Gregory:2010gx, Cassani:2011sv}.
Actually the present work is motivated by this paper where the authors have studied
the dynamics of a probe D-brane in the background \eqref{BacLif}. The action of the
D-brane probe is given by the DBI action.

It is important to mention  that due to lack of a rigorous realization of Lifshitz geometries in string theory, one can not treat the model as a top-down approach for inflationary model building. In particular we anticipate that as long as the string embedding
of Lifshitz geometries is absent, our model suffers from two important shortcomings when
compared to standard DBI inflation. First of all we  do not know how the model
allows to have the right
coupling between brane and the Ramond-Ramond (RR) $C_{(4)}$ potential which is necessary to get the correct slow-roll limit for the probe brane moving in the background. Secondly one does not know how to obtain the potential term for the inflaton field which is necessary to support inflation. In standard DBI inflation the potential term can originate from the couplings of the mobile D-brane to background RR fluxes, supplemented by any coupling of the D-brane to the compactification degrees of freedom, including quantum generated effects (see\cite{Silverstein:2003hf} for detail discussions on this point). Despite the above mentioned caveats, we consider our treatment as a phenomenological 
approach for inflationary model building with some motivations supported from string theory for the form of the potential or couplings required to support inflation.

The rest of paper is organized as follows. In section \ref{D3-back} we present the action 
of mobile brane in a Lifshitz background. In section \ref{background-cosmology} we promote this action to a cosmological set up and look into background inflation and the cosmological perturbations. In section \ref{dim-red} we obtain the general four-dimensional
action with arbitrary matter field and metric perturbations and speculate on its cosmological predictions.  The discussion  and conclusions are summarized in section \ref{summary}. The details of the gravitational dimensional reduction are relegated into
the Appendix.

%%%%%%%%%%%%%%%%%%%%%%%%%%%%%%%%%%%%%%%%%%%%%%%%%%

\section{D3-Branes in Lifshitz background}
\label{D3-back}

In this section we outline our setup of DBI inflation in Lifshitz background. Before we present our setup, we shall briefly review the conventional models of brane inflation, for a review of brane inflation see \cite{HenryTye:2006uv,Cline:2006hu,Burgess:2007pz,McAllister:2007bg,Baumann:2009ni, Mazumdar:2010sa} and references therein. We will heavily borrow the well-developed ideas and techniques in these models into our background. 

In conventional models of brane inflation \cite{dvali-tye,Alexander:2001ks,collection,Dvali:2001fw, Kachru:2003sx, Firouzjahi:2003zy, Burgess:2004kv, Buchel, Iizuka:2004ct, Firouzjahi:2005dh}
branes are moving in an AdS background \cite{Klebanov:2000hb, Giddings:2001yu, Dasgupta:1999ss}.
The action of the mobile branes is given by DBI action supplemented by the Chern-Simons term coming from the coupling of the mobile branes to the background fluxes. In the slow-roll models of warped brane inflation, the inflaton field is the distance between a pair of brane and anti-brane. The anti-brane is dynamically attracted towards the bottom of the AdS throat, the IR region,  where the mobile D3-brane is moving slowly towards it from the UV region. The inflationary potential is given by the mutual Coloumbic force between 
D3 and $\D$-branes. In this picture inflation ends when the distance between D3 and $\D$ becomes at the order of string scale $l_s$ where a tachyon develops in the open string spectrum and  the branes are annihilated. The energy released from the tensions of the colliding branes can be thought as the source of reheating. In the DBI  brane inflation \cite{Alishahiha:2004eh, Chen:2004gc, Shandera:2006ax, Bean:2007eh, Bean:2007hc, Chen:2006nt} the mobile brane is moving ultra relativistically inside the throat reaching the asymptotic speed limit. Due to non-standard form of kinetic energy from DBI action, the sound speed in cosmological perturbation theory can be much less than unity. This can have interesting observational consequences such as producing significant amount of non-Gaussianities \cite{Chen:2006nt}.

Here we would like to generalize the above picture of DBI inflation into Lifshitz background.
In light of the above picture of brane inflation, here are the key assumptions we make in our analysis. First, we assume that the Lifshitz background can actually be embedded in string theory.
Second, we assume that the concept of D3-branes with the right coupling to Ramond-Ramond 
four form potential $C_{(4)}$ actually exists in this picture 

As mentioned previously these are non-trivial assumptions which have yet to be justified in a rigorous string theory set up. Besides these two assumptions, there are other assumptions which are common in models of brane inflation. We assume that all complex structure moduli and the volume moduli are stabilized consistently. There can be back-reactions of volume modulus, background fluxes and branes to the mobile branes. In our effective field theory action, these are interpreted as corrections into the inflationary potential. As in conventional DBI inflation we shall take the phenomenological approach and parametrize the potential appropriately for the inflationary analysis. As usual in brane inflation, one can not avoid the issue of fine-tuning on the mass parameters or couplings  \cite{ Kachru:2003sx, Firouzjahi:2003zy, Burgess:2004kv,  Firouzjahi:2005dh, Baumann:2006th, Burgess:2006cb, Baumann:2007ah, Chen:2008au, Cline:2009pu, Hoi:2008gc}.
On these issues our position is the same as in conventional models of brane inflation.

%%%%%%%%%%%%%%%%%%%%%%%%%%%%%%%%%%%%%%%%%%%%%%%%%%
\subsection{The Action} 
In the Lifshitz background, the Lorentz symmetry in higher dimensions is broken with the metric in the following form 
\ba
\label{metr}
d s^2=-\left(\dfrac{r}{L} \right)^{2z} \md t^2+ \left(\dfrac{r}{L} \right)^2 \md \x^2+\left(\dfrac{L}{r} \right)^2 \md r^2+L^2 \md \Omega^2 \, .
\ea
Here $\md \Omega^2$ represents the metric in the angular directions which we left unspecified.
$r$ is the radial coordinate in the ``Lifshitz throat'' with the characteristic length scale  $L$. This setup is similar to AdS throat with $z=1$ where $L$ represents the AdS length scale of the throat. We keep the parameter $z$ undetermined taking value
in the range $  z \ge 1$\footnote{For the model considered in \cite{Kachru:2008yh} which
supports metrics of Lifshitz form the reality condition on the fluxes requires $z\ge 1$. Although in the present paper we do not consider a specific gravitational model admitting 
Lifshitz geometry, we still assume that this is the case too. } . 
All models of brane inflation so far focused on the AdS background with $z=1$.

As explained before, we assume that the mobile brane is moving 
in this background governed by the standard DBI and Chern-Simons actions. With this assumption, the action of the mobile D3 brane in background (\ref{metr}) is given by
\ba 
\label{action}
S=-T_3 \int \md^4 x  \left(\dfrac{r}{L} \right)^{3+z} \left(\sqrt{1-\left(\dfrac{L}{r}\right)^{2+2z}\dot{r}^2}-1 \right)  \, .
\ea
Here $T_3$ is the tension of the D3-brane and a dot here and below indicates derivative with respect to $t$. The first term in the bracket, containing the square root, is from the DBI part whereas the second term in the bracket originates from the Chern-Simons term. As before, this form of the action is motivated from the action of D3-brane in AdS throat where there is a specific relation between the background RR four form potential and the warp factor of the throat. This 
particular relation in AdS construction reflects the BPS or no-force condition of a mobile D3-brane in the background of AdS geometry in the limit where the brane is moving slowly.

We will be mostly interested in the DBI limit of the action (\ref{action}) where the brane
is moving ultra relativistically in the Lifshitz throat and one can not expand the square root
in Eq. (\ref{action}) perturbativly. However, it would be interesting to look into the slow-roll limit of Eq. (\ref{action})  when one can expand the square root perturbativly. In this limit one finds that the action reduces to that of free field with the action 
$S= \frac{1}{2} \int d^4 x \, \dot \phi^2$ where the normalized field $\phi$ is defined via
\ba
\label{dif-eq}
\dot{\phi} \equiv \sqrt{T_3} \, \dot{r}  \, \left(\dfrac{r}{L} \right)^{\frac{1-z}{2}}  \, .
\ea
We note that in the AdS limit with $z=1$ this gives the well-defined normalization that $\phi= \sqrt{T_3} r$. For general value of $z$ the above differential equation can be integrated easily to yield
\ba
\label{r-phi}
\phi =\begin{cases}
\mu_z \left(\dfrac{r}{L} \right)^{\frac{3-z}{2}}  & \text{if} \, \, z \neq 3
\\
\\
\mu_3 \ln(\dfrac{r}{L}) & \text{if}  \, \, z=3
\end{cases}  
\ea
where we have defined the parameters $\mu_z$ and $\mu_3$ via
\ba
\label{muz3}
\mu_3 \equiv \sqrt{T_3} L \quad \quad   , \quad \quad
\mu_z \equiv \dfrac{2 \mu_3}{3-z}   = \dfrac{2\sqrt{T_3}}{3-z}\, L \, .
\ea  
Note that for $z>3$, $\mu_z$ is negative so $\phi<0$. Also for the case of $z=3$, if
we start with $r<L$, then $\phi<0$. However, in our picture that the brane inside the throat is moving from the UV region towards the IR region, then $\phi$ is always decreasing for all value of $z$.  Also we see that the  relation between the physically normalized field $\phi$ and the 
radial coordinate $r$ is distinctly different for the case with scaling $z=3$. \footnote{In general $d$-dimensional space-time, the case $z=d-2$ is special such that $\phi \sim \ln r$ whereas for other values of $z\ge1$ the relation between $\phi$ and $r$ is a power law.}

Having obtained the physically normalized field $\phi$ in the slow-roll limit, the action \eqref{action} can be rewritten in more compact form which can be applicable to the general case when the brane is moving relativistically
\ba 
\label{DBI-Lif}
S=-\int \md^4 x  f^{-1} \left(\sqrt{1-f \dot{\phi}^2} -1\right)  \, ,
\ea 
where
\ba
f(\phi)=\begin{cases}
T_3^{-1} \left(\frac{\mu_z}{\phi} \right)^\alpha &        \text{if}  \, \, z \neq 3
\\
\\
T_3^{-1} e^{-\frac{6 \phi}{\mu_3}} & \text{if}  \, \, z=3
\end{cases}
\ea
and 
\ba
 \alpha \equiv \dfrac{2(3+z)}{3-z}  \, .
\ea
Note that in the case of AdS background we have $\alpha=4$ and $f \sim \phi^{-4}$
as expected.  
Interestingly, the form of the D3-action given in Eq. (\ref{DBI-Lif}) is identical to the standard form of DBI inflation. This will be a great help in performing the cosmological analysis in next sections and we can borrow many formulae from the standard DBI inflation
in AdS background. However, we note that despite this formal similarity, there are important
physical differences as one varies the scaling parameter  $z$.

%%%%%%%%%%%%%%%%%%%%%%%%%%%%%%%%%%%%%%%%%%%%%%%%%%
%%%%%%%%%%%%%%%%%%%%%%%%%%%%%%%%%%%%%%%%%%%%%%%%%%
\section{DBI Lifshitz Cosmology}
\label{background-cosmology}

Having presented our background, we promote it into a cosmological set up. We couple the world volume of the mobile branes to the FRW metric 
\ba
ds^2 = -dt^2 + a(t)^2 \md \x^2 \, ,
\ea
where $a(t)$ is the cosmological scale factor at the background isotropic and homogeneous level. In the cosmological background, the action of D3-brane in Eq. (\ref{action}) is transferred into
\ba 
\label{act}
S=-\int \md^4x \, a^3 \left[  f^{-1} \left(\sqrt{1-f \dot{\phi}^2}-1 \right)+V(\phi) \right] \, .
\ea
We have added the potential $V(\phi)$ needed for inflation by hand. As in standard brane inflation, there are many corrections to the dynamics of mobile branes in a throat. This includes the back-reactions from Kahler modulus, background fluxes or branes. Even for the case of 
a mobile brane in an AdS throat it is a non-trivial task to calculate all these corrections  from string theory \cite{Baumann:2006th, Burgess:2006cb, Baumann:2007ah, Chen:2008au}. At the phenomenological level, one may add the potential term to take into account these back-reactions. 
We follow the same phenomenological prescription here and absorb the unknown back-reactions into $V(\phi)$. Whether or not these back-reactions provide the right potential 
to sustain a long enough period of inflation is the infamous problem of fine-tuning in brane inflation  \cite{ Kachru:2003sx, Firouzjahi:2003zy, Burgess:2004kv,  Firouzjahi:2005dh, Baumann:2006th, Burgess:2006cb, Baumann:2007ah, Chen:2008au, Cline:2009pu, Hoi:2008gc}. Our position here is the same as in conventional models of brane inflation where 
it is assumed that the parameters of potential can be tuned, at least in principle, so one can obtain a successful period of inflation. 

The form of potential is not determined either. For the branes moving in a throat, one may 
expect a power law potential in term of its radial coordinate, $V \sim r^{n}$ with unknown 
number $n$, which can be either fractional or integer \cite{ Baumann:2007ah}. For our model where the brane is moving towards the IR region of the throat we take $n$ to be a positive number. For $z\neq3$, the scaling between $r$ and $\phi$ given in Eq. (\ref{r-phi})
yields $V \sim r^n \sim \phi^{2n/3-z}$, i.e. a power law inflation potential for $\phi$. On the other hand, for the case with $z=3$, Eq. (\ref{r-phi}) indicates that
$V \sim r^n \sim e^{ n \phi/ \mu_3 } $, i.e. an exponential potential for the inflaton field. 
These suggest that for our phenomenological investigations we can take the inflaton potential to
have the following form 
\ba
V(\phi)=\begin{cases}
V_0 \,  \left(  \frac{\phi}{\mu_z}   \right)^p &        \text{if}  \, \, z \neq 3
\\
V_0 \, e^{\frac{n \phi}{\mu_3}}  & \text{if}  \, \, z=3
\end{cases} 
\ea
where 
\ba
p\equiv \frac{2n}{3-z} \, .
\ea
The unknown parameter $n$ and the  energy scale $V_0$ are left undetermined and should be tuned to support long enough period of inflation and satisfy the WMAP constraints. 

In our phenomenological treatments, we have assumed that the form of potential is determined
as a function of $r$, such as $ V \sim r^n$ assumed above, and obtained the corresponding form of  $V(\phi)$. Instead, we could have chosen to start with a fixed phenomenological potential $V(\phi)$ and read off its corresponding form $V(r)$. However, the latter approach is less natural. In our setup with arbitrary value of $z$, the relation between $r$ and $\phi$ is
non-linear so if we start with $V(\phi)$ then the form of $V(r)$ would be completely different for the cases of $z=3$ and $z \neq 3$. The naturalness in starting with $V(r)$, as we employed above, originates from the fact that  to calculate $V(r)$ one has to incorporate the effects of background fluxes and volume moduli of string compactification. In this picture $r$, being a real string theory coordinate, is more  natural and the resulted potential from the back-reactions is expected to be a function of $r$ in the form of $V(r)$.

We  note that the case $z=1, n=2$ corresponds to conventional model of DBI inflation
with the potential $m^2 \phi^2/2$ which is vastly studied in the literature.  In the analysis below, we consider the cosmological predictions of our model for different values of $z$ and $n$.

%%%%%%%%%%%%%%%%%%%%%%%%%%%%%%%%%%%%%%%%%%%%%%%%%%
%%%%%%%%%%%%%%%%%%%%%%%%%%%%%%%%%%%%%%%%%%%%%%%%%%
\subsection{Background Cosmology }

We couple the action (\ref{act}) to the effective four-dimensional gravity and write down the background cosmological equations. 
In the analysis below we mostly follow  \cite{Shandera:2006ax} in notations and methods. 

As usual the Friedmann equation and the energy conservation equation are
\ba
\label{friedmann}
3 H^2 = \dfrac{\rho}{M_P^2}   \quad , \quad   \dot \rho + 3 H (\rho+p) =0 \, .
\ea
Here $H= \frac{\dot a}{a}$ is the Hubble expansion rate, $\rho$ and $p$ respectively are the energy density and the pressure
\ba
\label{p}
\rho = f^{-1} (\gamma -1)+ V
\quad , \quad 
p=f^{-1}( 1- \gamma^{-1}) -V \,  ,
\ea
in which $\gamma$ is the so called ``Lorentz factor" defined by
\ba
\gamma \equiv \dfrac{1}{\sqrt{1-f \dot{\phi}^2}}  \, .
\ea
The modified Klein-Gordon equation for the inflaton field is 
\ba
\label{phi-eq}
\ddot{\phi}+3H\gamma^{-2}\dot{\phi}+\frac{3}{2}\frac{{f'}}{f}\dot{\phi}^2-\frac{{f'}}{f^2}+\gamma^{-3} \left({V'}+\frac{{f'}}{f^2} \right)=0 \, .
\ea
Following \cite{Shandera:2006ax}, one can cast these equations into Hamilton-Jacobi forms which are more suitable for analytical purposes. Since $\phi$ is monotonically decreasing 
as time goes by, we can use $\phi$ as the clock and express the physical parameters in terms of
$\phi$. This yields 
\ba
\label{HJ}
3 M_P^2 H(\phi)^2 &=& V(\phi)+f^{-1}(\gamma(\phi) -1) \nonumber \\ 
\gamma(\phi) &=& \sqrt{1+4 M_P^4 f(\phi) H^{'}(\phi)^2}  \nonumber \\ 
\dot{\phi}(\phi) &=& \dfrac{-2 M_P^2 H'}{\gamma(\phi)}  
\ea 
where $H'= \partial H/\partial \phi$.

We are interested in the limit where the brane is moving ultra relativistically inside the throat with $\gamma \gg 1$. This corresponds to brane  ``speed limit''  where
$\dot \phi \simeq -\frac{1}{\sqrt{f}}$. For $z= 3$ this yields 
\ba
\label{speed-z=3}
\phi_{speed} \rightarrow - \frac{\mu_3}{3} \ln \left( \frac{3 t}{L}   \right)    \quad \quad 
(z= 3)
\ea
This indicates that the speed limit is reached only logarithmically in time, much slower than  the case of AdS background with $z=1$ where  $\phi_{speed} \sim \frac{1}{t}$ \cite{Silverstein:2003hf}. Note that  the brane is moving towards the IR region where $r<L$ so that $\phi_{speed} <0$ as seen above. 

On the other hand, for $z\neq3$, we obtain 
\ba
\label{speed-z}
\phi_{speed} \rightarrow  \mu_z \left( \frac{z t}{L} 
\right)^{(z-3)/2z}  \quad \quad 
(z\neq 3)
\ea
In particular, with $z=1$ we obtain $\phi_{speed} \sim \frac{1}{t}$ as expected for 
standard DBI inflation.  Note that for $z>3$, $\phi$ and $\mu_z$ are negative so as time goes by 
$| \phi|$ increases. This is indicated by the fact that the speed limit is speeding up with
a positive power of $t$. On the other hand, for $1 \leq z \leq 3$, $\phi$ is positive so
a negative power of $t$ above indicates that $\phi \rightarrow 0^+$. Of course this picture will
terminate if we want to have a graceful exit from inflation. In our picture, we assume that
there is an anti-brane at the bottom of the throat, $r=r_0$. Once the distance between the mobile brane and anti-brane reaches at the order of string length scale then the system becomes tachyonic and inflation ends quickly after brane and anti-brane annihilation. So the speed limit
described above is in  the idealistic limit where the Columbic force between brane and anti-brane is neglected and the brane is moving indefinitely towards the bottom of the throat in the absence of anti-branes.

It is also instructive to look into the number of e-folding, $N$. To solve the flatness and the horizon problem we assume $N =60$. 
Using $d N = H dt$, one obtains
\ba
\label{N-eq}
N = \int_{\phi_f}^{\phi_i}  d \phi \, H(\phi) \sqrt {f(\phi)} \, ,
\ea
where $\phi_i$ and $\phi_f$ respectively represent the initial and the final value of the inflaton field.
In the speed limit one can easily integrate this expression and find $N$ as a function of $\phi_i$ and $\phi_f$.
However, it turns out that the results can be expressed more easily in terms of $r$-coordinate
%For  $z\neq \frac{n}{2}$ we have
%\ba \label{N-r} N\simeq  \sqrt{\frac{4 V_0 \mu_3^2}{3 M_P^2 T_3}} \frac{1}{n-2z} \left( \frac{r}{L} \right)^{\frac{n-2z}{2}} \huge {|}_{r_f}^{r_i} \,  \quad \quad  \left(z\neq \frac{n}{2} \right) \ea
%whereas for $z=\frac{n}{2}$, one obtains 
%\ba \label{N-r-z=2n} N\simeq  \sqrt{\frac{4 V_0 \mu_3^2}{3 M_P^2 T_3}} \ln \left( \frac{r_i}{r_f}  \right) \quad \quad  \left( z=\frac{n}{2} \right) \ea
\ba
\label{N-r}
N=\sqrt{\frac{ V_0 \mu_3^2}{3 M_P^2 T_3}}
\begin{cases}
  \frac{2}{n-2z} \left( \frac{r}{L}
\right)^{\frac{n-2z}{2}} \huge {|}_{r_f}^{r_i}  &          \left( z \neq \frac{n}{2}\right)
\\
\ln \left( \frac{r_i}{r_f}
\right)  & \left( z=\frac{n}{2} \right)
\end{cases} 
\ea
where $r_i$ and $r_f$, respectively, are the initial and the final values of the mobile brane's  coordinates.

Here we pause to address the issue of ending inflation where $N$ is an increasing function of $r_f$ as in the case of $n<2z$ in Eq. (\ref{N-r}). Interestingly this situation indicates that inflation, or a dS solution, is attractor towards the IR region of the throat. As the brane is moving towards the IR region, inflation proceeds indefinitely. However,  as mentioned above, inflation ends in this situations when the distance between brane and anti-brane, located at  the bottom of throat $r=r_0$,  reaches at the order of string scale. Using the metric (\ref{metr}) the physical  distance $d$ between the brane and anti-brane is 
$d=  L \ln \left(r_f/r_0\right)$ where $r_f$ is the final value of the brane position. Setting
$d=l_s$ for $l_s$ being the string length scale results in 
\ba
\label{rf}
r_f = r_0 \exp \left( \frac{l_s}{L}
\right) \, .
\ea

For the case $z=3$, this results in
\ba
\label{phif-z=3}
\phi_f = \mu_3 \left[ \frac{l_s}{L} - \ln \left( \frac{L}{r_0}\right)  \right] \, .
\ea
Depending on the sign of the term inside the bracket, $\phi_f$ can be either positive or negative.
However, one can easily arrange such that $L/r_0$ is exponentially large as in GKP 
construction \cite{Giddings:2001yu} such that $\phi_f$ is typically expected to be negative.
On the other hand, for $z\neq 3$ the value of $\phi_f$ is obtained to be
\ba
\label{phif-z}
\phi_f = \mu_z \exp \left[ \frac{3-z}{2}  \left(  \frac{l_s}{L} - \ln \left( \frac{L}{r_0}\right)      \right) \right] \, .
\ea 
As expected, for $L/r_0$ exponentially large, $\phi_f \rightarrow 0^+$ for $1\leq z <3$
whereas for $z>3$ $\phi_f$ decrease towards more negative values.

After reviewing the background cosmology, we turn to perturbations in general DBI Lifshitz inflation.

%%%%%%%%%%%%%%%%%%%%%%%%%%%%%%%%%%%%%%%%%%%%%%%%
%%%%%%%%%%%%%%%%%%%%%%%%%%%%%%%%%%%%%%%%%%%%%%%%

\section{Dimensional Reduction and Cosmological Perturbations}
\label{dim-red}

Having presented our background inflationary solution, it is time to address the question of cosmological perturbations in this model. As we shall see below, the perturbations in this model is drastically different than the standard DBI inflation model. The reason is that after dimensional reduction from 5D into 4D, the general covariance is lost explicitly in 4D.
The loss of general covariance is explicit both in the field theory sector and in the gravitational sector. This is because we start from a theory in higher dimension which breaks Lorentz invariance explicitly when $z \neq 1$.  Having this said, one may worry that the lack of 4D general covariance may destroy the effective FRW cosmology which we have obtained at the homogeneous and isotropic level. Here we verify that this is not a problem for the background cosmology. To see this, let us uplift metric (\ref{BacLif}) into a homogeneous and isotropic FRW background
 \ba
 \label{FRW-Lif}
d s^2=-N(t)^2 \left(\dfrac{r}{L} \right)^{2z} \md t^2+ \left(\dfrac{r}{L} \right)^2  a(t)^2 \md \x^2+\left(\dfrac{L}{r} \right)^2 \md r^2
 \ea
 where as usual $N(t)$ is the lapse function added in order to find the Friedmann constraint equation. Calculating the five-dimensional Ricci scalar $^{(5)}R$ from metric (\ref{FRW-Lif}) we have
 \ba
 \label{R5-1}
 ^{(5)}R =   \left( \frac{L}{r}\right)^{2 z} R_{FRW} -\frac{1}{L^2} \left( 12+ 6 z + 2 z^2
 \right)
 \ea
 where $R_{FRW}$ is the four-dimensional Ricci scalar constructed from the four-dimensional FRW metric
 \ba
 R_{FRW} \equiv \frac{6}{N^2} \left(\frac{\ddot a}{d} + (\frac{\dot a}{a})^2 - \frac{\dot a}{a} \frac{\dot N}{N}  \right) \, .
 \ea
Note that  the last term in bracket in Eq. (\ref{R5-1}) has no dynamics in terms of the four-dimensional metric and contributes only to the effective cosmological constant which can be canceled by similar terms from other fields (such as the massive gauge fields which we do not consider here). As a result, the four-dimensional gravitational action, $^{(4)}S_{gr}$, is
\ba
\label{Sgr}
^{(4)}S_{gr} \subset \frac{1}{2 \kappa^2}\int d^5 x \sqrt{- G}\,  ^{(5)}R  =
\frac{M_P^2}{2}
\int d^4 x N a(t)^3   R_{FRW}  
%\left(   \frac{1}{2 \kappa^2}   \int d r \left(\frac{r}{L}\right)^{2-z} \right) 
\ea 
 with the identification
 \ba
 \label{Mp}
 M_P^2 \equiv   \frac{1}{ \kappa^2}   \int_V d r \left(\frac{r}{L}\right)^{2-z} \, ,
 \ea
 where $\kappa$ is the five-dimensional gravitational coupling and the integration over the compact volume $V$ is supposed to be finite in order to obtain a finite four-dimensional gravitational coupling. As usual, this can be achieved by gluing the Lifshitz throat into the bulk of CY compactification. Eq. (\ref{Sgr}) clearly demonstrates that we recover the standard FRW action for the isotropic and homogeneous cosmology and our results for the background cosmology coupled with the matter sector given in Eq. (\ref{act}) is indeed justified.

Now we would like to uplift metric (\ref{BacLif}) into a generic cosmological background where the effective 4D metric, once the  extra $r$-coordinate is integrated out, 
is given by $g_{\mu \nu}= \{ g_{00}, g_{0i}, g_{ij} \}$.  The natural ansatz is 
\ba
\label{Lif-cosmo}
d s^2=g_{00}\left(\dfrac{r}{L} \right)^{2z} \md t^2+ g_{ij}\left(\dfrac{r}{L} \right)^2 
\md x^i \md x^j+ 2 g_{0i} \left(\dfrac{r}{L} \right)^{\kappa} \md t\,  \md x^i +
\left(\dfrac{L}{r} \right)^2 \md r^2
\ea
Here $g_{00}, g_{ij}$ and $g_{0i}$ are arbitrary functions of 4D space-time coordinates $x^{\mu}$. We have left the arbitrary scaling parameter $\kappa$ in the $0i$ component of the metric. However, demanding that the scaling  (\ref{scaling}) still  to hold requires that $\kappa = z+1$.

Before we proceed with the perturbations it should be stressed that the ansatz (\ref{Lif-cosmo}) 
may not be a solution of the five-dimensional Einstein equation. In principle there are other fields, such as a massive gauge field, which should be added into the action in order to support the Lifshitz geometry (\ref{BacLif}). To study the perturbations one should also study the perturbations of these non-gravitational fields. 
In our analysis below we shall concentrate only on the gravitational sector, given by the usual Einstein-Hilbert term $\sqrt{-|G_{MN}|} {^{(5)}}R$, and the inflaton perturbations. 
The ansatz (\ref{Lif-cosmo}) with only the metric and inflaton perturbations are rich enough to demonstrate the nature of perturbations in our analysis.

 %%%%%%%%%%%%%%%%%%%%%%%%%%%%%%%%%%%%%%%%%%%%%%%%%
 
\subsection{Matter Action}

Now we obtain the matter sector Lagrangian with the metric perturbations given in Eq. (\ref{Lif-cosmo}). The Lagrangian for the matter sector is obtained by considering a probe brane moving inside the Lifshitz background whose embedding $X^M$ is a general function of 4D 
space-time coordinates
\ba
X^M = \left( x^\mu, r( x^\nu)  \right)
\ea
where $r( x^\nu) $  indicates the brane position inside the throat. 

The induced metric on the brane, $\bar g_{\mu \nu}$, is given by
\ba
\bar g_{\mu \nu} = \frac{\partial X^M}{\partial{x^\mu} }  \frac{\partial X^M}{\partial{x^\mu} }
G_{MN} \, ,
\ea 
where $G_{MN}$ is the 5D background metric given by Eq. (\ref{BacLif}).  One obtains
\ba
\bar g_{00} &=& g_{00} \left(\frac{r}{L}\right)^{2z} + \left(\frac{L}{r}\right)^2 \dot r^2 
\nonumber\\
\bar g_{ij} &=& g_{ij} \left(\frac{r}{L}\right)^2 + \left(\frac{L}{r}\right)^2 \partial_i r \partial_j r
\nonumber\\
 \bar g_{0i} &=& g_{0i} \left(\frac{r}{L}\right)^{z+1} + \left(\frac{L}{r}\right)^2 \dot r \, \partial_i r  \, .
\ea
To obtain the matter sector action  we are interested in calculating $\sqrt{- | \bar g_{\mu \nu} |}$. After some long calculations one obtains
 \ba
 \label{general-metric}
 | \bar g_{\mu \nu} | = | g_{\mu \nu} |  \left(\frac{r}{L} \right)^{z+3}
 \left[ 1+ g^{00} \left(\frac{r}{L}\right)^{-2(z+1)} \dot r^2
 +  \left(\frac{r}{L}\right)^{-4} g^{ij} \partial_i r \partial_j r + 2 \left(\frac{r}{L}\right)^{-(z+3)}  g^{0i} \dot r \, \partial_i r
 \right] \, .
 \ea
Note that here $g^{\mu \nu}$ is the inverse of metric $g_{\mu \nu}$ in the usual sense. 
This equation has a very interesting structure. We note that in the limit where $z=1$, the 4D theory becomes general covariant in the matter sector  as expected. However, for arbitrary $z \neq1$ the general covariance is lost 
in the field theory. In order to connect it to our background inflation analysis, we introduce the physical field $\phi$ as before
\ba
\label{dif-eq}
\dot{\phi} \equiv \sqrt{T_3} \, \dot{r}  \, \left(\dfrac{r}{L} \right)^{\frac{1-z}{2}}  \, .
\ea
Working with $\phi(x^\alpha)$, the DBI action becomes
\ba
-T_3 \sqrt{- | \bar g_{\mu \nu} |} = -\sqrt{- | g_{\mu \nu} |} f(\phi)^{-1} \left[1+ 
f(\phi) g^{00}  \left(\partial_t \phi\right)^2 + h(\phi) g^{i j} \partial_i  \phi \partial_j \phi
+ 2 \ell(\phi) g^{0i} \partial_t \phi \partial_i \phi
\right]^{1/2}
\ea
 in which for the case $z\neq 3$ we have 
 \ba
 f(\phi) \equiv T_3^{-1} \left(\frac{\mu_z}{\phi} \right)^\alpha \quad , \quad
 h(\phi)\equiv T_3^{-1} \left(\frac{\mu_z}{\phi} \right)^{\alpha'} \quad  \quad  z\neq 3
 \ea
with 
 \ba
 \alpha \equiv \dfrac{2(3+z)}{3-z}  \quad , \quad 
 \alpha' \equiv \dfrac{2(5-z)}{3-z} \, .
\ea
On the other hand, for $z=3$ one has
\ba
f(\phi)\equiv T_3^{-1} e^{-\frac{6 \phi}{\mu_3}} \quad , \quad 
h(\phi) \equiv T_3^{-1} e^{-\frac{2 \phi}{\mu_3}} \quad , \quad  z=3
\ea 
 Very interestingly, in both cases one has $\ell(\phi) = \sqrt{f(\phi) h(\phi)}$. 

Now adding the Chern-Simons part and demanding that the theory reduces to a free field theory in the slow-roll limit the total brane action is 
\ba
\label{D3-action}
{\cal{L}}_{D3} =  -\sqrt{- | g_{\mu \nu} |} f(\phi)^{-1} \left\{ \left[1+ 
f(\phi) g^{00}  \left(\partial_t \phi\right)^2 + h(\phi) g^{i j} \partial_i  \phi \partial_j \phi
+ 2 \ell(\phi) g^{0i} \partial_t \phi \partial_i \phi
\right]^{1/2} -1\right\} \, .
\ea 
 This is an interesting theory. It clearly shows that the general covariance is lost in the matter sector. One can restore the general covariance only when $z=1$ and $f(\phi)= h(\phi)=\ell(\phi)$. 
 
 Now that the matter sector breaks the general covariance explicitly, we expect that the general covariance is also lost in 4D gravitational theory. 
 
 %%%%%%%%%%%%%%%%%%%%%%%%%%%%%%%%%%%%%%%%%%%%%%%%%
 \subsection{Gravitational Action}
 \label{gr-action}
 
In this subsection we would like to calculate the four dimensional gravitational action  
$\sqrt{- |g_{\mu \nu}|}\,  ^{(4)}R$ from the five-dimensional gravitational action $\sqrt{- G}\,  ^{(5)}R$ where $G_{MN}$ is given  by Eq. (\ref{Lif-cosmo}).   To start, from Eq. (\ref{Lif-cosmo}) we have  $\sqrt {-|G_{MN}|} = \sqrt{-|g_{\mu \nu}|} \left( \frac{r}{L} \right)^{z+2}   $ . It is interesting  that the 4D and 5D determinants are proportional to each other.  One can check that this is possible if one takes $\kappa = z+1$
 as we did. 
 
Now we consider the gravitational dimensional reduction. One key observation is that although 
the four-dimensional diffeomorphism is expected to be broken explicitly but  the metric (\ref{Lif-cosmo}) is still invariant under  the three-dimensional diffeomorphism  and time rescaling given by
 \ba
 \label{3d-diff}
 t \rightarrow \tilde t = \tilde t(t) \quad , \quad
 x^i \rightarrow \tilde x^i= \tilde x^i (x^j, t) \, .
 \ea
As a result the dimensionally reduced gravitational action is expected to be written in such a way that the three-dimensional diffeomorphism invariance is explicit.  In general one expects that the result to be expressed in terms of the intrinsic three-dimensional Ricci scalar $^{(3)}R$ and the extrinsic curvature  $K_{ij}$ which describes the curvature of the three-dimensional surface $t=constant$ in four dimension. 

To find $K_{ij}$ it is better as usual to employ the ADM formalism where the metric 
 (\ref{Lif-cosmo}) is written as
 \ba
 \label{ADM}
 ds^2 = -(\frac{r}{L})^{2z} N^2 dt^2 + g_{ij} \left[ (\frac{r}{L}) dx^i + N^i (\frac{r}{L})^z dt
 \right]  \left[ (\frac{r}{L}) dx^j + N^j (\frac{r}{L})^z dt
 \right]  \, ,
 \ea
 so
  \ba
 g_{00} = -N^2 + g_{ij} N^i N^j  \quad , \quad g_{ij} = ^{(3)}g_{ij} \quad ,\quad
 g_{0i} = g_{ij} N^j \, .
 \ea
 Note that we use $G_{MN}$ for the 5D metric while $g_{\mu \nu}$
is the 4D metric as defined in  (\ref{Lif-cosmo}). One can also check that 
\ba
g^{00} = -\frac{1}{N^2} \quad , \quad g^{ij} = ^{(3)} g^{ij} - \frac{N^i N^j}{N^2}
\quad , \quad g^{0i} = \frac{N^i}{N^2}
\ea
It is important to note that we consider a general metric perturbations so 
$N=N(t, x, y, z)$, i.e. a ``non-projectable'' metric \cite{Blas:2009qj} which is still consistent
with the time rescaling and the three-dimensional diffeomorphism given in Eq. (\ref{3d-diff}).

The extrinsic curvature is given by
\ba
K_{ij} = \frac{1}{2N} \left(  \dot g_{ij} - ^{(3)} {\nabla_i N_j} - ^{(3)} \nabla_j N_i 
\right)
\ea
where $^{(3)} \nabla$ represents the three-dimensional covariant derivative constructed from 
the metric $^{(3)} g_{ij}$.

We relegate the details of the gravitational dimensional reduction analysis into 
Appendix \ref{gdr} and quote the final results. After neglecting terms which are either 
total derivatives or contribute only to the effective cosmological constant term such as the last term
in Eq. (\ref{R5-1}) one obtains
\ba
\label{gravity1}
\sqrt {-|G_{MN}|} \, {^{(5)}}R \rightarrow \left(\frac{r}{L}\right)^{2-z} \sqrt {-|g_{\mu\nu}|}
\left[ {^{(4)}}R + \Omega\,   {^{(3)}}R  
\right] \, 
\ea
where 
 \ba
 \label{Omega}
 \Omega\equiv  (\frac{r}{L})^{2(z-1)} -1  \, .
 \ea
In the limit where $z=1$ and $\Omega=0$ we recover the standard four dimensional gravity
which is general covariant. However, for non-zero value of $\Omega$ the 4D general covariance is broken into three-dimensional general covariance. This is also in light with the loss of the 4D general covariance in the matter sector, Eq. (\ref{D3-action}). Note that for the FRW background with ${^{(3)}}R=0$, Eq. (\ref{gravity1}) reproduces  Eq. (\ref{R5-1}) as expected. On the other hand noting that $^{(4)}R = ^{(3)}R +K^2 - K_{ij}K^{ij}$, 
we can replace $^{(3)}R$ in terms of the extrinsic curvature to obtain
\ba
\label{gravity2}
\sqrt {-|G_{MN}|}\,  {^{(5)}}R \rightarrow \left(\frac{r}{L}\right)^{2-z} \sqrt {-|g_{\mu\nu}|}
\left[ \left(1+ \Omega \right)  {^{(4)}}R + \Omega \left( K^2 - K_{ij} K^{ij}       \right)
\right]
\ea

Now we perform the compactification. Performing the integral over the internal manifold volume $V$ one obtains
\ba
\label{grav-final}
\frac{1}{2 \kappa^2}  \int d^5 x \sqrt{-G} ^{(5)}R &&\rightarrow  \int d^4 x \, \sqrt{-|g_{\mu \nu}|}\frac{M_P^2}{2} \left[ ^{(4)}R + \bar \Omega\,  {^{(3)}}R  \right] \nonumber\\
&&= \int d^4 x \, \sqrt{-|g_{\mu \nu}|}\frac{M_P^2}{2} \left[  (1+ \bar \Omega) ^{(4)}R + \bar \Omega\,  
 \left( K^2 - K_{ij} K^{ij}   \right)\right] \, ,
\ea
where $M_P$ is defined as in Eq. (\ref{Mp})
and
\ba
\bar \Omega  \equiv \frac{1}{\kappa^2 M_P^2} \int_V dr \, \Omega \, \left(\frac{r}{L}\right)^{2-z}  = \frac{1}{\kappa^2 M_P^2} \int_V dr \,  \left( \frac{r}{L} \right)^{2-z}
\left[  -1 + \left( \frac{r}{L} \right)^{2(z-1)}
\right] \, .
\ea
Eq. (\ref{grav-final}) is our final result for the gravitational dimensional reduction. 

One natural question is the magnitude and the sign of the parameter $\bar \Omega$ which measures the extent of the 4D general covariance breaking. To estimate $\bar \Omega$ one has to know the details of the compactification. As in conventional models of brane inflation suppose the Lifshitz throat is extended between the radial coordinate $r_0 < r<R$ so 
$r_0$ is the IR end of the throat while $R$ is the UV region where the Lifshitz throat is assumed to be glued smoothly to the bulk of the CY compactification. Using Eq. (\ref{Mp})
to eliminate $\kappa^2 M_P^2$ we obtain
\ba
\label{omeg-1}
\bar \Omega  = -1 + \frac{\int_V dr \left(\frac{r}{L}\right)^z}{\int_V dr\,  \left(\frac{r}{L}\right)^{2-z}} \, .
\ea
To calculate $\bar \Omega$ we consider the cases $z<3$,  $z=3$ and $z>3$
separately.
For $z < 3$ and assuming that $r_0 \ll R$, we obtain
%\ba \bar \Omega \simeq  -1  + \frac{3-z}{z+1} \left(\frac{R}{L}\right)^{2(z-1)}  \left[ 1- 
% \left(\frac{r_0}{R}\right)^{3-z} \right]   \quad \quad z \neq 3 \ea 
\ba
\bar \Omega \simeq -1 + \frac{3-z}{z+1} \left(\frac{R}{L}\right)^{2(z-1)} 
\quad \quad z < 3   \, .
\ea
Depending on the value of the ratio $R/L$, $\bar \Omega$ can be either positive or negative.
Defining $R_c$ as when $\bar \Omega=0$, we obtain 
\ba
R_c = L  \left(\frac{3-z}{z+1}\right)^{1/2 (z-1)} \, .
\ea
So for $R<R_c (R>R_c)$ one has $\bar \Omega <0 (\bar\Omega >0)$.
Interestingly at $R=R_c$, $\bar \Omega$ vanishes although we have $z \neq 1$. This means that the  4D theory is general covariant in the gravitational sector if the Lifshitz throat is glued to the CY compactification at the critical radius $R=R_c$. For generic value of $R$ and assuming that $R/L$ typically is  order of few, we expect that the magnitude of $\bar \Omega$ to be order of few.

On the other hand for $z>3$, we have 
\ba
\bar \Omega \simeq -1 + \frac{z-3}{z+1} \left(\frac{R}{L}\right)^{2(z-1)} 
 \left(\frac{r_0}{R}\right)^{z-3} 
\quad \quad z > 3   \, .
\ea
We know that $r_0/R$ is exponentially small \cite{Giddings:2001yu}, so for values of 
$R \lesssim L$ we expect to have $\bar \Omega <0$. However, by choosing $R$ to be 
much bigger than $L$ one can tune $\bar \Omega$ to cross zero and become positive too. 

Finally for $z=3$, we obtain
\ba
\bar \Omega \simeq -1 +  \frac{(\frac{R}{L})^4}{4 \ln(\frac{R}{r_0})} \,  
\quad \quad \quad \quad z = 3   \, .
\ea
As for the case
$z<3$, one can find a critical value $R_c$ where $\bar \Omega=0$. Assuming that typically
$ \ln R/r_0 $ is order few, we conclude that $R_c \sim L$. 
For $R < R_c (R>R_c)$ we have $\bar \Omega <0 (\bar \Omega >0)$.

Having obtained the action for the matter and the gravitational sectors, Eqs. (\ref{D3-action})  and (\ref{grav-final}),  the total action, $S= \int d^4 x { L}_{total}$, therefore is 
\ba
\label{S-total}
{ L}_{total} &=& \sqrt{-|g_{\mu \nu}|}\frac{M_P^2}{2} \left[ ^{(4)}R + \bar \Omega\,  {^{(3)}}R  \right] \nonumber\\
 &-&\sqrt{- | g_{\mu \nu} |} f(\phi)^{-1} \left\{ \left[1+ 
f(\phi) g^{00}  \left(\partial_t \phi\right)^2 + h(\phi) g^{i j} \partial_i  \phi \partial_j \phi
+ 2 \ell(\phi) g^{0i} \partial_t \phi \partial_i \phi
\right]^{1/2} -1\right\} 
\ea
Eq. (\ref{S-total}) indicates an interesting theory where the general covariance is broken explicitly both in the matter and the gravitational sectors. This action is similar to the classifications employed in effective field theory of inflation (EFTI) literature \cite{Cheung:2007st}. In EFTI studies the 4D general covariance is broken because the inflaton field 
introduces preferred three-dimensional time foliation slices.
This is also the case for our matter sector where the non-trivial $\phi$-dependences break explicitly the 4D general covariance. However, the lack of 4D covariance in gravitational sector is more non-trivial. It originates from the fact that the higher-dimensional 
theory breaks the Lorentz invariance at the level of solution. After compactification, this manifests itself with an extra scalar field degree of freedom. This is reminiscent of the extra scalar degree of freedom in Horava-Lifshitz theory of gravity \cite{Horava:2009uw, Charmousis:2009tc, Li:2009bg, Blas:2009yd, Blas:2009qj}.

Having presented our total action in Eq. (\ref{S-total}) one can look into the cosmological perturbation theory in details and check the observational predictions. Specifically the model seems to have interesting predictions for non-Gaussianities.  Note that already in standard DBI inflation with $\bar \Omega=0$ one has significant non-Gaussianities. Now due to explicit breakdown of four-dimensional diffeomorphism 
one naturally expects to obtain non-Gaussianities with different shapes. The 
situation here is to some extent similar  to the EFTI literature on non-Gaussianities, 
for example see   \cite{Cheung:2007st, Senatore:2009gt,  Bartolo:2010bj,  Bartolo:2010di, Baumann:2011su}
and references therein.  We would like to come back to the questions of cosmological perturbations and non-Gaussianities in a future publication. 

%%%%%%%%%%%%%%%%%%%%%%%%%%%%%%%%%%%%%%%%%%%%%%%%%
\section{Summary and Discussions}
\label{summary}

In this paper a new model of DBI inflation is presented where the mobile brane is moving ultra relativistically inside a Lifshitz background with arbitrary scaling exponent $z$. This is a generalization of standard DBI inflation inside an AdS throat with $z=1$.

This work is mainly at the phenomenological level. As expressed previously, we do not have a rigorous construction of our Lifshitz throat within string theory set up. It is an open question as how one can embed the Lifshitz background rigorously in string theory. Furthermore, the concept of brane with the right RR coupling is not well-understood in this set up. We have employed the phenomenological approach that the Lifshitz throat with mobile branes, as required for brane inflation, can be embedded in principle in string theory. Furthermore, as in standard brane inflation, we assumed that
one can obtain the inflationary potential from the back-reactions of the mobile branes with the background fluxes and volume modulus. Whether or not the potential has the right form and the right couplings to support long enough period of inflation  is the question of fine-tuning and on this issue our model is on the same footing as conventional models of brane inflation.

In this work we have not studied the slow-roll brane inflation. This is because the potential between brane and anti-brane required for slow-roll inflation, as in \cite{Kachru:2003sx}, is not understood for arbitrary value of $z$ in Lifshitz background.  It would be interesting to calculate the inter-brane potential for generic value of $z$ so both slow-roll and fast-roll brane inflation can be combined for a more richer  inflationary model building in this set up.

After dimensional reduction we found that the four-dimensional general covariance is broken explicitly both in the matter and the gravitational sectors. The reason is that we start from a theory where the Lorentz invariance is broken at the level of solution. Performing the dimensional reduction to four dimension, this manifests itself in parameter $\bar \Omega$ which measures the level of 4D diffeomorphism breaking. Depending on how Lifshitz throat is glued to the bulk of CY compactification, we find that $\bar \Omega$ can have either signs and can be order of few.  The total action, including the general metric and matter perturbations,  is obtained in Eq. (\ref{S-total}). One expects the model has interesting cosmological predictions, specially for the magnitude and the shapes of the non-Gaussianities. We would like to come back to these question in a future publication.

%%%%%%%%%%%%%%%%%%%%%%%%%%%%%%%%%%%%%%%%%%%%%%%%%%
%%%%%%%%%%%%%%%%%%%%%%%%%%%%%%%%%%%%%%%%%%%%%%%%%%
\section*{Acknowledgement}

We would like to thank  P. Creminelli, R. Fareghbal, J. Maldacena, 
A. Nicolis, L. Senatore and E. Silverstein for useful discussions. We also thank
M. M. Sheikh-Jabbari and the anonymous JCAP referee for the comments on perturbations in
the earlier version of the draft. We specially thank Kazuya Koyama for many insightful discussions  and comments. H.F. would like to thank ICG for the hospitality during the revision of this work.

%%%%%%%%%%%%%%%%%%%%%%%%%%%%%%%%%%%%%%%%%%%%%%%%%%

%%%%%%%%%%%%%%%%%%%%%%%%%%%%%%%%%%%%%%%%%%%%%%%%%
\appendix

\section{Gravitational Dimensional Reduction}
\label{gdr}

In this appendix we provide the details of the gravitational dimensional reduction.  As mentioned in subsection \ref{gr-action}, the metric (\ref{Lif-cosmo}) is invariant under 
three-dimensional time-dependent diffeomorphism defined in Eq. (\ref{3d-diff}). As a result we expect the dimensionally reduced gravitational action to be written in terms of geometric quantities such as $^{(3)} R$ and $K_{ij}$ which respect this symmetry. Furthermore, due to 
the three-dimensional diffeomorphism invariance we can go to a coordinate system (gauge) where $g_{0 \mu}=0$ which simplifies the calculations considerably. After obtaining the results in this gauge, we express them in terms of three-dimensional covariant quantity such as $K_{ij}$ which ensure the validity of the results independent of the coordinate system.  

In the gauge where $g_{0\mu}=0$ one can check that 
\ba
\label{RMN}
^{(5)}R_{00} &=&^{(4)} R_{00} - \Omega
 \left[ -\partial_i  ^{(4)}\Gamma^i_{00} + 2 ^{(4)}\Gamma^i_{00}
{^{(4)}\Gamma^0_{0i}} -^{(4)}\Gamma^i_{00} \left( ^{(4)}\Gamma^0_{0i} + ^{(4)}\Gamma^j_{ij} \right)
\right] - (\frac{r}{L})^{2z} \frac{z^2+3z}{L^2} g_{00}
 \nonumber\\
^{(5)}R_{ij} &=& ^{(4)}R_{ij} + \left(\frac{r}{L}\right)^{-2(z-1)} \Omega
 \left[ -\partial_0 ^{(4)}\Gamma^0_{ij} +  ^{(4)}\Gamma^0_{il} {^{(4)}\Gamma^l_{0j} }+
 ^{(4)}\Gamma^l_{0i} {^{(4)}\Gamma^0_{lj}}
-^{(4)}\Gamma^0_{ij}( ^{(4)}\Gamma^0_{00} +^{(4)} \Gamma^l_{0l} )
\right]  \nonumber\\ &-& (\frac{r}{L})^2  \, \frac{z+3}{L^2}g_{ij}
 \nonumber\\
^{(5)}R_{55} &=& -\frac{3+z^2}{r^2}
\ea
Here $^{(4)}\Gamma^\mu_{\nu \kappa}$ is calculated from the four-dimensional metric $g_{\mu \nu}$ and  $\Omega$ is given via Eq. (\ref{Omega})
 
Note that in the limit where $z=1$, $\Omega$ vanishes. As we shall see below
 $\Omega$ measures the degree of the lack of the  four-dimensional diffeomorphism invariance.

With some efforts one can check that
\ba
^{(5)}R = \left(\frac{r}{L}\right)^{-2z} \left[ {^{(4)}}R  + \Omega \left( g^{ij} {^{(4)}R_{ij}}
- a_0 +a_1 \right) \right]  - \frac{1}{L^2} \left( 12+ 6 z + 2 z^2 \right) 
\ea
where 
 \ba
 a_0 &\equiv& g^{00}  \left[ -\partial_i {^{(4)}}\Gamma^i_{00} + 2 {^{(4)}}\Gamma^i_{00}
{^{(4)}}\Gamma^0_{0i} -{^{(4)}}\Gamma^i_{00}( {^{(4)}}\Gamma^0_{0i} + {^{(4)}}\Gamma^j_{ij} )
\right]  \nonumber\\
a_1 &\equiv& g^{ij}  \left[ -\partial_0 {^{(4)}}\Gamma^0_{ij} +  {^{(4)}}\Gamma^0_{il} {^{(4)}}\Gamma^l_{0j} +
 {^{(4)}}\Gamma^l_{0i}{^{(4)}}\Gamma^0_{lj}
-{^{(4)}}\Gamma^0_{ij}( {^{(4)}}\Gamma^0_{00} + {^{(4)}}\Gamma^l_{0l} )
\right]
 \ea
 
 To get better insight about the forms of different terms above, we go to the ADM formalism defined in Eq. (\ref{ADM}). With the gauge $g_{0\mu}=0$ we have $N^i=0$ while 
 $N$ and $g_{ij}$ are arbitrary functions of space and time. Calculating $a_0$ explicitly one can check that
 \ba
a_0= \frac{1}{N} \nabla^2 N
 \ea
where $\nabla^2$ is constructed from $g_{ij}$.  On the other hands, many terms in  $a_1$ and 
$^{(4)}R_{ij}$ cancel out and one obtains
\ba
g^{ij} {^{(4)}R_{ij}} +a_1 = {^{(3)}}R + \frac{1}{N} \nabla^2 N \, .
\ea
Combining all terms together, we obtain
\ba
\label{dim-red1}
^{(5)}R =  \left(\frac{r}{L}\right)^{-2z} \left[ {^{(4)}}R + \Omega \left( {^{(3)}}R +
 \frac{2}{N} \nabla^2 N \right) \right] - \frac{1}{L^2} \left( 12+ 6 z + 2 z^2 \right) 
\ea
Considering the addition factor of $N$  coming from 
$\sqrt {-|G_{MN}|} ^{(5)}R = N \sqrt{|g_{ij}|} \left( \frac{r}{L} \right)^{z+2}{ ^{(5)}R}$ the term 
$\nabla^2 N/N$ is transformed into  a total derivative term integrated over three-dimensional space which can be discarded. Also neglecting the last term in Eq. (\ref{dim-red1}) which contributes into the effective cosmological constant and will be canceled by the contributions from other field in the bulk, yield 
\ba
\label{dim-red2}
\sqrt {-|G_{MN}|} \, {^{(5)}}R \rightarrow \left(\frac{r}{L}\right)^{2-z} \sqrt {-|g_{\mu\nu}|}
\left[ {^{(4)}}R + \Omega\,   {^{(3)}}R  
\right]
\ea
This is our final result for the gravitational dimensional reduction. Note that in the limit where 
$\Omega=0$, one recovers the standard four-dimensional gravitational theory which is
general covariant. However, with a non-zero value of $\Omega$, the general covariance is broken in 4D. Also note that for the  FRW background, Eq. (\ref{dim-red2}) reduces to 
Eq. (\ref{R5-1}) as expected.

Alternatively, one can use the Gauss-Codazzi equation 
$^{(4)}R = ^{(3)}R -K^2 + K_{ij}K^{ij}$ to eliminate ${^{(3)}}R $ in terms of 
extrinsic curvature and obtain
\ba
\label{dim-red3}
\sqrt {-|G_{MN}|}\,  {^{(5)}}R \rightarrow \left(\frac{r}{L}\right)^{2-z} \sqrt {-|g_{\mu\nu}|}
\left[ \left(1+ \Omega \right)  {^{(4)}}R + \Omega \left( K^2 - K_{ij} K^{ij}       \right)
\right]
\ea

%%%%%%%%%%%%%%%%%%%%%%%%%%%%%%%%%%%%%%%%%%%%%%%%

\end{document}